\documentclass[12pt,epsf,amstex]{article}
\usepackage [dvips]{graphicx}
\usepackage{amsmath}
\usepackage{amssymb}
\usepackage{epsfig}

\addtocounter{secnumdepth}{1}
\begin{document}

\newcommand{\br}{\bar{r}}
\newcommand{\bbeta}{\bar{\beta}}
\newcommand{\bgamma}{\bar{\gamma}}
\newcommand{\bR}{{\bf{R}}}
\newcommand{\bS}{{\bf{S}}}
\newcommand{\half}{\frac{1}{2}}
\newcommand{\thalf}{\tfrac{1}{2}}
\newcommand{\summ}{\sum_{m=1}^n}
\newcommand{\sumqno}{\sum_{q\neq 0}}
\newcommand{\tsum}{\Sigma}
\newcommand{\bsA}{\mathbf{A}}
\newcommand{\bsV}{\mathbf{V}}
\newcommand{\bsE}{\mathbf{E}}
\newcommand{\bsZ}{\hat{\mathbf{Z}}}
\newcommand{\bse}{\mbox{\bf{1}}}

\newcommand{\dd}{\mbox{d}}
\newcommand{\ee}{\mbox{e}}
\newcommand{\p}{\partial}
\renewcommand{\i}{\rm i}

\newcommand{\bn}{\bar{n}}
\newcommand{\bN}{\bar{N}}
\newcommand{\cL}{\cal L}
\newcommand{\cW}{\cal W}

\newcommand{\la}{\langle}
\newcommand{\ra}{\rangle}

\newcommand{\beq}{\begin{equation}}
\newcommand{\eeq}{\end{equation}}
\newcommand{\bea}{\begin{eqnarray}}
\newcommand{\eea}{\end{eqnarray}}

\newcommand{\erf}{\mbox{erf}}
\newcommand{\PG}{P^{\rm Gauss}}
\newcommand{\pih}{\pi^{\half}}
\newcommand{\emttrho}{(1-\frac{2}{3}\rho)}

\def\lsim{\:\raisebox{-0.5ex}{$\stackrel{\textstyle<}{\sim}$}\:}
\def\gsim{\:\raisebox{-0.5ex}{$\stackrel{\textstyle>}{\sim}$}\:}

\numberwithin{equation}{section}

\thispagestyle{empty}
\title{{\Large {\bf 
A note on $q$-Gaussians and non-Gaussians\\[2mm]
in statistical mechanics\\
\phantom{xxx} }}}

\author{{ H.\,J. Hilhorst and G.\,Schehr}\\[3mm]
{\small Laboratoire de Physique Th\'eorique\,$^1$,
B\^atiment 210, Universit\'e de Paris-Sud}\\[-1mm]
{\small 91405 Orsay Cedex, France}\\}

\maketitle
\begin{small}
\begin{abstract}
\noindent 

The sum of $N$ sufficiently strongly correlated
random variables will not in general be Gaussian distributed 
in the limit $N\to\infty$.
We revisit examples of sums $x$ that
have recently been put forward as instances 
of variables obeying a $q$-Gaussian law,
that is, one of type $\mbox{cst}\times[1-(1-q)x^2]^{1/(1-q)}$.
We show by explicit calculation that the probability distributions
in the examples are actually analytically different from $q$-Gaussians, 
in spite of numerically resembling them very closely. 
Although $q$-Gaussians exhibit many interesting properties,
the examples investigated do not support the idea that
they play a special role as limit distributions
of correlated sums.\\

\end{abstract}
\end{small}
\vspace{45mm}

\noindent LPT Orsay 07/27\\
{\small $^1$Laboratoire associ\'e au Centre National de la
Recherche Scientifique - UMR 8627}
\newpage


\section{Introduction} 
\label{secintroduction}
\vspace{5mm}

The central limit theorem says that
the sum of $N$ sufficiently weakly correlated 
random variables, 
when properly scaled with $N$ and in the limit $N\to\infty$, 
has a Gaussian probability distribution.
In many places in physics, however,
strongly correlated variables occur. Causes of strong correlations may be,
for example, the presence of long-range interaction or 
memory effects.
On the basis of a generalization of the statistical mechanical formalism
\cite{Tsallis88} it has been suggested 
that in certain strongly correlated
systems such many-variable sums
are distributed according to a $q$-Gaussian law.
This law is a generalization of the ordinary Gaussian and arises,
in particular, when one maximizes the generalized entropy 
which is at the basis of the theory. 
Pursuant to this idea
$q$-Gaussians have been advanced to describe 
velocity distributions
in a Hamiltonian mean-field model (HMF) of classical rotators
\cite{LRT02}, in turbulent
Couette-Taylor flow \cite{BLS01}, and in cellular
aggregates \cite{URGS01}; and were also proposed
for the velocity distribution of galaxy clusters \cite{LKetal96}
and the temperature fluctuations 
in the cosmic microwave background \cite{BTV06}.
In at least the case of the HMF, divergent interpretations of Monte Carlo
results have led to vivid discussions in the literature
(see {\it e.g.,} \cite{BO06,TRPB07,YBD07,BD05}).
\vspace{2mm}

For a single variable $x$ the
$q$-Gaussian probability distribution $G_q(x)$ is given by
\beq
G_q(x) = b^\half C_q \Big[1\,-\,\frac{1-q}{3-q}\,{b} x^2 \Big]^{\frac{1}{1-q}},
\label{defGq}
\eeq
where $C_q$ is a normalization constant and the scale factor $b$
controls the width.
For $-\infty<q<1$, which will be the range of $q$ values of interest to us, 
$G_q(x)$ is bell-shaped and has its
support restricted to the interval $[-x_{\rm m},x_{\rm m}]$,
where
$x_{\rm m}=\big( \frac{1-q}{3-q}\,b \big)^{-\half}$.
The denominator $3-q$ in (\ref{defGq}) is there
merely to guarantee that for $q\to -\infty$ the support of the 
$q$-Gaussian remains finite. The special case
$G_{-\infty}$ is the block function equal to $\half b^\half$ on 
$[-b^{-\half},b^{-\half}]$.
For $q\to 1$ expression (\ref{defGq}) reduces to the ordinary 
Gaussian $G_1(x)=(b/2\pi)^\half \exp(-\half b x^2)$.

Independently of their applicability in statistical mechanics, 
$q$-Gaussians have several interesting mathematical properties.
For example, the multivariate $q$-Gaussian 
[for which $x^2$ in (\ref{defGq}) is replaced with a symmetric bilinear 
form $\sum_{\mu\nu} x_\mu A_{\mu\nu} x_\nu$]
has marginal distributions that are
$q'$-Gaussians with $q'\neq q$ (see \cite{UT07}). 
\vspace{2mm}

We will be concerned specifically with the question of whether
$q$-Gaussians can emerge as probability distributions of 
{\it sums of correlated random variables}. 
Attempts in the literature to demonstrate how 
such $q$-Gaussian distributions may arise, 
have consisted of constructing and numerically analyzing
simplified model systems. 
A successful model should not only explain the origin
of the $q$-Gaussian, but also express  
the parameter $q$ in terms of more common concepts such as 
interactions and correlations.
These model studies leave for later investigation the question
of whether the properties preassigned to the model may actually 
be realized in physical systems.

In this note we will examine two model systems taken 
from the recent literature \cite{TMNT06,MTGM06}.
Both feature sums that were put
forward as candidates for being $q$-Gaussian distributed. 
We show by explicit calculation that in both examples
the probability distributions
are actually analytically different from $q$-Gaussians, 
even though very closely resembling them numerically. 
Notwithstanding the interesting properties exhibited by $q$-Gaussians,
the two examples studied therefore do not lend support to the idea that
these functions play a role as limit distributions
of correlated sums.


\section{First example}
\label{secfirstcase}


\subsection{A set of strongly correlated variables}
\label{secdefu1uN}

Thistleton {\it et al.} \cite{TMNT06} consider what is probably the simplest
imaginable instance of a strongly correlated system. It consists of
$N$ identically distributed variables $u_1,u_2,$ $\ldots,u_N$
such that the probability distribution $P_1$
of each individual $u_j$ is the block function
\beq
P_1(u_j)=1, 
\qquad (-\tfrac{1}{2} \leq u_j \leq \tfrac{1}{2})\,,
\label{exprP1Uj}
\eeq
and $P_1(u_j)=0$ elsewhere.
The $u_j$ have, furthermore, strong mean-field type correlations, 
as expressed by their covariance matrix 
\beq
\la u_ju_k \ra =
\left\{
\begin{array}{ll}
\tfrac{1}{12} & \quad (j=k),\\[2mm]
r             & \quad (j \neq k).
\end{array}
\right.
\label{expravUjUk}
\eeq
where $0 <r \leq 1$.
The authors of reference \cite{TMNT06}
are then interested in finding the probability distribution 
${\cal P}_N(U)$ of the scaled sum 
\beq
U = N^{-1}\sum_{j=1}^N u_j
\label{defU}
\eeq
and in obtaining its limit ${\cal P}(U)=\lim_{N\to\infty}{\cal P}_N(U)$.
It will turn out in this mean-field problem that in order for
the limit $N\to\infty$ to exist, 
the appropriate scaling is the one with $N^{-1}$ given in
(\ref{defU}) rather than the conventional one with $N^{-1/2}$
applicable to independent variables. It is therefore clear
that we cannot expect the limits $r\to 0$ and $N\to\infty$ to commute.

Determining ${\cal P}_N(U)$ requires the knowledge of
the full joint probability distribution 
$P_N(u_1, u_2,\ldots,u_N)$
of the $u_j$.
This distribution is fixed
implicitly by the way the $u_j$ are constructed \cite{TMNT06}, namely, 
from a Gaussian distribution that one knows how to generate numerically.
Let the $N$ correlated Gaussian variables $z_1,z_2,\ldots,z_N$ 
be distributed according to
\beq
\PG_N(z_1,z_2,\ldots,z_N) = \big[ (2\pi)^N \det M \big]^{-\tfrac{1}{2}}
\exp\Big(\! -\tfrac{1}{2}\sum_{j=1}^N\sum_{k=1}^N z_j M^{-1}_{jk}z_k \Big).
\label{defpz1zN}
\eeq
Then the $z_j$ have the covariance matrix $\la z_jz_k\ra=M_{jk}$,
for which will take 
\beq
M_{jk} =
\left\{
\begin{array}{ll}
1       & \quad (j=k),\\[2mm]
\rho    & \quad (j \neq k).
\end{array}
\right.
\label{exprMjk}
\eeq
where $\rho\in(0,1]$ is a parameter. 
It follows that the elements of the inverse matrix $M^{-1}$ are
\beq
M^{-1}_{jk} =
\left\{
\begin{array}{rl}
\alpha &  \quad (j=k),\\[2mm]
-\beta &  \quad (j \neq k),
\end{array}
\right.
\label{exprMm1jk}
\eeq
with
\beq
\alpha=\frac{1+(N-2)\rho}{(1-\rho)[1+(N-1)\rho]}\,,\qquad
\beta =\frac{\rho}       {(1-\rho)[1+(N-1)\rho]}\,.
\label{expralphabeta}
\eeq
The marginal distribution $\PG_1(z_j)$ of $z_j$ 
derived from (\ref{defpz1zN})
is a Gaussian of unit variance.
The variables $u_j$ are now constructed \cite{TMNT06} from the $z_j$
according to
\bea
u_j &=& \int_0^{z_j}\!\dd x\, \PG_1(x) \nonumber\\[2mm]
    &=& \frac{1}{2}\, \erf \Big( \frac{z_j}{\sqrt{2}} \Big),
        \qquad (j=1,\ldots,N).
\label{defuj}
\eea
Then $u_j$ has the block distribution (\ref{exprP1Uj}) and is
correlated to the other $u_k$ according to (\ref{expravUjUk}) with an $r$
that may be determined as a function of $\rho$.

On the basis of numerical evidence
Thistleton {\it et al.} \cite{TMNT06}
conjectured that the probability distribution ${\cal P}(U)$ 
of the scaled sum $U$ of these variables was a $q$-Gaussian with an index $q$
related to $\rho$ according to 
\beq
q = \frac{1-\tfrac{5}{3}\rho}{1-\rho}\,.
\label{relqrho1}
\eeq
This example is sufficiently simple to allow for an exact analytic treatment,
which we will present now.


\subsection{The limiting law ${\cal P}(U)$}
\label{seclimitinglaw1}

In this subsection we replace the block function $P_1$ considered
in (\ref{exprP1Uj}) by an arbitrary symmetric function $P_1(u)$. 
Let there be $N$ correlated random variables
$u_j$, $j=1,2,\ldots,N$ having an arbitrary symmetric marginal distribution
$P_1(u_j)$ of unit variance and having correlations
$\la u_ju_k \ra=r$ for $j \neq k$.
Variables $u_j$ having these properties 
may be obtained from the $z_j$ by
\beq
\int_0^{z_j}\!\dd x\, \PG_1(x) = \int_0^{u_j}\!\dd x\, P_1(x),
\qquad (j=1,\ldots,N),
\label{defgenuj}
\eeq
which generalizes (\ref{defuj}) and which we will abbreviate as
\beq
u_j = h(z_j), \qquad (j=1,\ldots,N).
\label{defhz}
\eeq
The expression for the law ${\cal P}_N(U)$ 
follows directly from its definition,
\bea
{\cal P}_N(U) &=& \big[ (2\pi)^N \det M \big]^{-\frac{1}{2}}
\int_{-\infty}^{\infty}\!\dd z_1\ldots\dd z_N\,
\delta\big( U-N^{-1}\sum_ih(z_i) \big) \nonumber\\[2mm]
&& \times
\exp \Big( -\tfrac{1}{2}\sum_j \sum_k z_j M^{-1}_{jk}z_k \Big).
\label{exprinitPU}
\eea
The quadratic form in the exponential in (\ref{exprinitPU})
simplifies according to
\beq
\sum_j \sum_k z_j M^{-1}_{jk}z_k \,=\, (\alpha+\beta)\sum_jz_j^2
-\beta\Big( \sum_jz_j \Big)^2.
\label{simplified}
\eeq
Upon substituting (\ref{simplified}) in (\ref{exprinitPU}) and introducing
integral representations for $\exp\!\big(\tfrac{1}{2}\beta(\sum_jz_j)^2\big)$
and for the delta function, we 
find that the integrations on the $z_j$ factorize and we get
\bea
{\cal P}_N(U) &=& \big[N (2\pi)^N\det M \big]^{-\frac{1}{2}}
\int_{-\infty}^{\infty}\frac{\dd\nu}{\sqrt{2\pi}}\,\ee^{-\frac{1}{2}N\nu^2}
\int_{-\infty}^{\infty}\frac{\dd\lambda}{2\pi}\,\ee^{-{\i}\lambda U}
\nonumber\\[2mm]
&& \times\,
\Big( \int_{-\infty}^\infty \!\dd z\, \ee^
{-\half(\alpha+\beta)z^2 + \nu(N\beta)^{1/2}z + 
{\i}\lambda N^{-1}h(z)} \Big)^N.
\label{exprPU2}
\eea
We use that $\alpha + \beta = 1/(1-\rho)$
and shift from $z$ to a new variable of integration
\beq
w = (1-\rho)^{-\half}\big( z-\nu(1-\rho)(N\beta)^\half \big).
\label{defw}
\eeq
Completing the square in the exponential in (\ref{exprPU2})
produces a factor 
$\exp\!\big( \tfrac{N\beta}{2}\times(1-\rho)\nu^2 \big)$ 
which may be taken outside the integral and whose
$N$th power combines with $\exp(-\half N\nu^2)$ to yield
$\exp(-\half\gamma\nu^2)$ where $\gamma=N(1-\rho)/[1+(N-1)\rho]$.
The asymptotic $N$ dependence of
this coefficient, namely $\gamma\simeq(1-\rho)/\rho$,
is the result of a cancellation of terms of order $N$.
Using that $\det M = [1+(N-1)\rho](1-\rho)^{N-1}$ 
and rearranging some of the prefactors we transform (\ref{exprPU2}) into
\bea
{\cal P}_N(U) &=& \Big( \frac{(1-\rho)N}{2\pi[1+(N-1)\rho]} 
                \Big)^\half
\int_{-\infty}^\infty\!\dd\nu\, \ee^{-\gamma\nu^2} 
\int_{-\infty}^{\infty}\frac{\dd\lambda}{2\pi}\,\ee^{-{\i}\lambda U}
\nonumber\\[2mm]
&&\times\,
\Big( \frac{1}{\sqrt{2\pi}} \int_{-\infty}^\infty\! \dd w\, 
      \ee^{-\half w^2 + {\i}\lambda N^{-1}h((w+\nu)\sqrt{1-\rho})} 
     \Big)^N.
\label{exprPU3}
\eea
At this point we perform an asymptotic expansion in powers of $N^{-1}$
of the right hand side of equation (\ref{exprPU3}).
This gives
\bea
{\cal P}(U) &=& \lim_{N\to\infty} \Big( \frac{1-\rho}{2\pi\rho} \Big)^\half
\int_{-\infty}^\infty\!\dd\nu\, \ee^{-\frac{1-\rho}{2\rho}\nu^2}
\int_{-\infty}^{\infty}\frac{\dd\lambda}{2\pi}\,\ee^{-{\i}\lambda U}
\nonumber\\[2mm]
\times\,
\Big( 1 &+& \frac{{\i\lambda}}{N} \int_{-\infty}^\infty\! 
\frac{\dd w}{\sqrt{2\pi}}\, 
      \ee^{-\half w^2}\,h\big( (w+\nu)\sqrt{1-\rho} \big) + {\cal O}(N^{-2})
     \Big)^N.
\label{exprPU4}
\eea
We now define
\beq
k(\nu) = U - \int_{-\infty}^\infty\! \frac{\dd w}{\sqrt{2\pi}}\, 
      \ee^{-\half w^2}\,h\big( (w+\nu)\sqrt{1-\rho} \big)
\label{defknu}
\eeq
and simplify (\ref{exprPU4}) further to
\bea
{\cal P}(U) &=& \Big( \frac{1-\rho}{2\pi\rho} \Big)^\half
\int_{-\infty}^\infty\!\dd\nu\, \ee^{-\frac{1-\rho}{2\rho}\nu^2}
\int_{-\infty}^{\infty}\frac{\dd\lambda}{2\pi}\,
\ee^{-{\i}(\lambda k(\nu)}
\nonumber\\[2mm]
&=& \Big( \frac{1-\rho}{2\pi\rho} \Big)^\half
\int_{-\infty}^\infty\!\dd\nu\, \ee^{-\frac{1-\rho}{2\rho}\nu^2}
\delta\big(k(\nu)\big).
\label{exprPU5}
\eea
Let $\nu_*(U)$ be the unique solution of
\beq
k(\nu_*) = 0,
\label{defnustar}
\eeq
so that we may write 
$\delta\big(k(\nu)\big)=\delta(\nu-\nu_*)|k^\prime(\nu_*)|^{-1}$.
Then, in the general case under consideration here, 
our final result for the limit function ${\cal P}(U)$ follows from
(\ref{exprPU5}) as 
\beq
{\cal P}(U) = \Big( \frac{1-\rho}{2\pi\rho} \Big)^\half
\big| k^\prime\big(\nu_*(U)\big) \big|^{-1}
\,\exp\Big(\! -\frac{1-\rho}{2\rho}\big[ \nu_*(U) \big]^2 \Big), \qquad
(-\tfrac{1}{2} \leq U \leq \tfrac{1}{2}),
\label{resfingenPU}
\eeq
in which $k$ is defined by (\ref{defknu}), $\nu_*$ by (\ref{defnustar}),
and we have set $k^\prime=\dd k/\dd\nu$.


\subsection{Special case of a block function}
\label{secspecial}

The limit function ${\cal P}(U)$
given by equations (\ref{resfingenPU}), (\ref{defknu}), and (\ref{defnustar})
can be found fully explicitly for the 
special case of the block function distribution (\ref{exprP1Uj})
studied by Thistleton {\it et al.}~\cite{TMNT06}.
According to equations (\ref{defuj}) and (\ref{defhz}), 
we have in that case $h(z)=\half\,\erf(z/\sqrt{2})$, which
when substituted in (\ref{defknu}) yields
\bea
k^\prime(\nu) &=& \pi^{-\half}\kappa\,\ee^{-\kappa^2\nu^2}, \qquad
\kappa \equiv \Big( \frac{1-\rho}{2(2-\rho)} \Big)^\half,
\nonumber\\[2mm]
k(\nu) &=& U - \thalf\,\erf(\kappa\nu),
\nonumber\\[2mm]
\nu_* &=& \kappa^{-1}\erf^{-1}(2U),
\label{knuspecial}
\eea
where\, $\erf^{-1}$\, denotes the inverse error function.
After straightforward substitution of (\ref{knuspecial})
in the general expression (\ref{resfingenPU})
we obtain 
\beq
{\cal P}(U) = \Big( \frac{2-\rho}{\rho} \Big)^\half
         \exp \Big(\! -\frac{2(1-\rho)}{\rho} 
                      \big[ \,\erf^{-1}(2U)\, \big]^2\, \Big),
         \qquad (-\tfrac{1}{2} \leq U \leq \tfrac{1}{2}).
\label{resfinPU}
\eeq
This is our end result for
the distribution ${\cal P}(U)$ of the scaled sum $U$
in the special case studied by Thistleton {\it et al.}~\cite{TMNT06}.
It is not the $q$-Gaussian that the authors speculated it to be.

\begin{figure}
\begin{center}
\scalebox{.60}
{\includegraphics{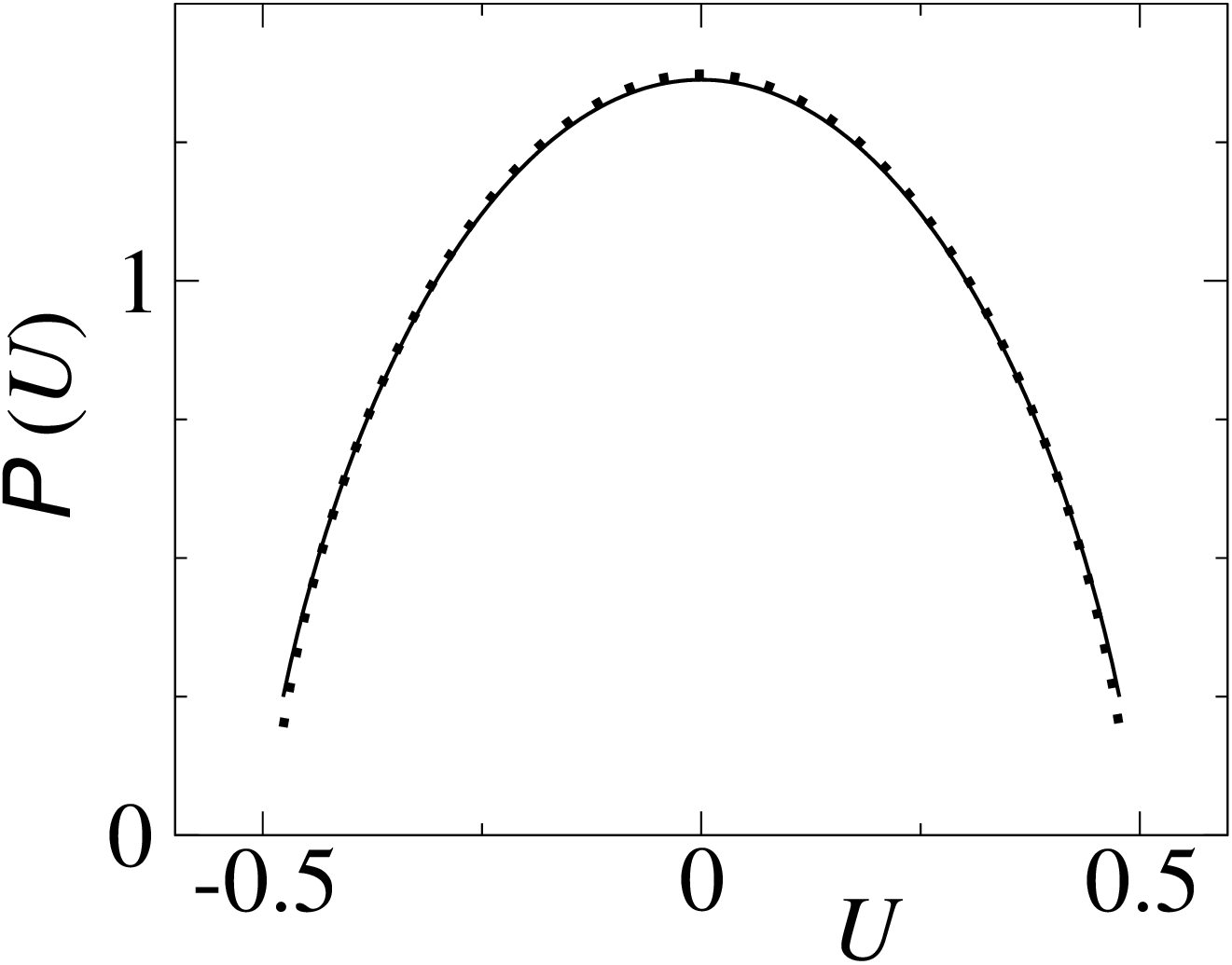}}
\end{center}
\caption{{\small Solid line: the exact limit function ${\cal P}(U)$ 
given by equation (\ref{resfinPU}) for the case $\rho=\tfrac{7}{10}$. 
Dotted line: the $q$-Gaussian 
approximation to this function for $q=-\tfrac{5}{9}$ (see text).
The difference between the two curves is of the order of the thickness of the
dotted line and just barely visible to the eye.
}} 
\label{figurePU}
\end{figure}

\subsection{Discussion of ${\cal P}(U)$}
\label{secdiscussion1}

For $U\to\pm\half$ we have that
$\erf^{-1}(2U)\to\pm\infty$ and therefore
${\cal P}(U)=0$ outside the interval $[-\half,\half]$.
One may verify explicitly the normalization 
$\int_{-1/2}^{1/2} {\cal P}(U)\dd U = 1$.
We consider now successively the limits $\rho\to 0$ and $\rho\to 1$.
The inverse error function has the expansion \cite{Wolfram}
\beq
\erf^{-1}(2U) = \pih U + \mbox{$\frac{1}{3}$} (\pih U)^3 
+ \mbox{$\frac{7}{30}$} (\pih U)^5
+ \mbox{$\frac{127}{360}$} (\pih U)^7 
+ \mbox{$\frac{4369}{22\,680}$} (\pih U)^9
+ \ldots 
\label{expinverf}
\eeq
For $\rho\ll 1$ it suffices to keep the first term in this expansion
and substitution in (\ref{resfinPU}) shows that in that limit
${\cal P}(U)$ approaches a Gaussian of variance $\rho/(4\pi)$.
This Gaussian limit is nevertheless somewhat curious:
whereas it is true that for $\rho\to 0$ the $N$ variables become independent,
the usual Gaussian distribution
associated with their sum would require a scaling with
$N^{-1/2}$ instead of the present one with $N^{-1}$. 
Hence we have here the explicit demonstration of the non-commutativity of the
limits $\rho\to 0$ and $N\to\infty$.
 
In the other limit, $\rho\to 1$, one sees directly that
${\cal P}(U)$ tends to unity on the interval
$[-\half,\half]$. 
This second limit is heuristically obvious since for $\rho=1$
the random variables are fully correlated.
For $0<\rho<1$ the function ${\cal P}(U)$ interpolates between 
a Gaussian and a block function.
Since the $q$-Gaussian does the same for $1>q>-\infty$, 
one may ask, for given $\rho$, whether there is a corresponding
$q$ that gives the best fit. This will be done in the next subsection.


\subsection{Best $q$-Gaussian fit to ${\cal P}(U)$}
\label{secbestfit1}

We try to find the best $q$-Gaussian fit to 
solution (\ref{resfinPU}) by adjusting the values 
of $q$ and the width ${b}$ in (\ref{defGq}).
There is no unique way of defining a `best fit'. 
We will base ourselves on comparing the small-argument expansions of
${\cal P}(U)$ and $G_q(x)$. The one of $G_q(x)$ is immediate,
\beq
G_q(x)= b^\half C_q\, 
\exp\Big[ - \sum_{n=1}^\infty 
\frac{(1-q)^{n-1}}{n(3-q)^n} {b}^n x^{2n} \Big].
\label{expGqx}
\eeq
The expansion of ${\cal P}(U)$ is obtained by substituting (\ref{expinverf})
in (\ref{resfinPU}) and going through the algebra. One finds
\bea
{\cal P}(U) &=& \Big( \frac{2-\rho}{\rho} \Big)^\half
\exp\Big[ -\sum_{n=1}^\infty C_{2n} U^{2n} \Big]
\nonumber\\[2mm]
&=& \Big( \frac{2-\rho}{\rho} \Big)^\half
\exp\Big[ -\frac{2(1-\rho)}{\rho} \sum_{n=1}^\infty c_{2n} \pi^n U^{2n} \Big],
\label{expPU}
\eea
in which the coefficients $C_{2n}$ are defined by the $c_{2n}$
which in turn are given by
\beq
c_2 = 1,\quad
c_4 = \mbox{$\frac{2}{3}$},\quad
c_6 = \mbox{$\frac{26}{45}$},\quad
c_8 = \mbox{$\frac{176}{315}$},\quad
c_{10} = \mbox{$\frac{8138}{14\,175}$}\,,\ldots 
\label{coeffcn}
\eeq
The $q$-Gaussian (\ref{expGqx}) contains two adjustable parameters, 
$q$ and ${b}$.
For the best fit we will adopt the criterion that
the coefficients of the quadratic and quartic terms
in the expansions (\ref{expGqx}) and (\ref{expPU}) coincide.
This leads directly to
the relation $C_4/C_2^2 = \half (1-q)$. Working out the remaining
algebra using (\ref{expGqx}), (\ref{expPU}),
and (\ref{coeffcn}), we obtain
\beq
q = \frac{1- \frac{5}{3}\rho }{1-\rho}\,, \qquad 
{b} = \frac{4\pi(1-\tfrac{2}{3}\rho)}{\rho}\,.
\label{fitbq}
\eeq
The first one of these equations is precisely the result (\ref{relqrho1}) 
conjectured by
Thistleton {\it et al.}~\cite{TMNT06} on the basis of their numerical
simulations! 
As an example, we show in figure \ref{figurePU} 
the exact solution (\ref{resfinPU})
for $\rho=\frac{7}{10}$ (solid line) together with its
$q$-Gaussian approximation (dotted line), which according to (\ref{fitbq}) 
has $q=-\frac{5}{9}$. The difference between the two is of the order of the
thickness of the dotted line and just barely visible to the eye.

To get an analytic idea of the quality of the fit (\ref{fitbq})
we consider  
the ratio of the $q$-Gaussian
approximation to the true solution.
>From (\ref{expGqx})-(\ref{fitbq}) we deduce that
\beq
\frac{G_q(U)}{{\cal P}(U)} =
b^\half C_q\Big( \frac{\rho}{2-\rho} \Big)^\half\,
\exp\Big[ -\frac{2(1-\rho)}{\rho}
\sum_{n=3}^\infty \epsilon_{2n}\pi^nU^{2n} \Big],
\label{ratioGP}
\eeq
where we set
\beq
\epsilon_{2n}=n^{-1}\big( \tfrac{4}{3} \big)^{n-1}-c_{2n}, 
\qquad (n=1,2,\ldots).
\label{defepsilon}
\eeq
By construction $\epsilon_2=\epsilon_4=0$.
It is remarkable that for the next few coefficients
the terms in the difference (\ref{defepsilon}) largely cancel. 
Evaluation gives, in particular, 
\beq
\epsilon_6    = \mbox{$\frac{2}{135} = 0.01481$}, \quad 
\epsilon_8    = \mbox{$\frac{32}{945} = 0.03386$}, \quad 
\epsilon_{10} = \mbox{$\frac{822}{14\,175} = 0.05799$}.  
\label{coeffepsilonn}
\eeq
The smallness of these coefficients explains analytically why the  
$q$-Gaussian fit (\ref{fitbq}) is so excellent.

Finally, we observe that with the criterion employed here
the two functions ${\cal P}$ and $G_q$ do not exactly coincide for $U=0$.
Using the explicit expression for $C_q$ 
we find for their ratio in the origin
\bea
\frac{G_q(0)}{{\cal P}(0)} \,=\, 
b^\half C_q\Big( \frac{\rho}{2-\rho} \Big)^\half
&=& \frac{3(2-\rho)}{2(3-2\rho)}\,
\frac{\big[ \Gamma(\frac{3}{2\rho}-\half) \big]^2}
{\Gamma(\frac{3}{2\rho}-1)\Gamma(\frac{3}{2\rho})}
\nonumber\\[2mm]
&=& 1 - \mbox{$\frac{1}{144}$}\rho^2 + \ldots,
\label{ratioGUPor}
\eea
where the last line represents a small-$\rho$ expansion.
Again we see that the deviation from unity is very small.
Hence our analysis confirms the high quality of the fit.


\subsection{Remark on the general case (\ref{resfingenPU}) }
\label{secfinalremark}

The general case that we set out to consider in section \ref{seclimitinglaw1} 
is specified by an arbitrary marginal one-variable distribution $P_1(x)$.
The corresponding limit distribution ${\cal P}(U)$
found in (\ref{resfingenPU}) is then obtained in terms of a function $\nu_*(U)$
that follows from $P_1(X)$ via relations (\ref{defgenuj}), (\ref{defhz}), 
(\ref{defknu}), and (\ref{defnustar}).
Although ${\cal P}(U)$ is not a $q$-Gaussian
in the special case where $P_1(x)$ is a block function,
it is likely that an appropriate choice of $P_1(x)$ will lead to a
$\nu_*(U)$ for which ${\cal P}(U)$ is a $q$-Gaussian.
The conditions for this to happen constitute a rather complicated set of
equations that we will not study here.
The important point is that in the same way not only a $q$-Gaussian but
{\it any\,} limit function ${\cal P}(U)$ out of a large class can be obtained;
there is no indication that $q$-Gaussians play a special role.


\section{Second example: the MTG sum}
\label{secMTG}


\subsection{A special probability law}
\label{secdefinition2}

In an earlier attempt to construct a  
$q$-Gaussian distributed sum of variables, 
Moyano, Tsallis, and Gell-Mann (MTG) \cite{MTGM06} 
proposed the expression
\beq
R_{N}(n) = \sum_{i=N-n}^N(-1)^{i+n-N}\binom{n}{i+n-N}
\frac {p} { [i-(i-1)\,p^{\,\rho}]^{\frac{1}{\rho}} }\,,
\label{defrNn}
\eeq 
which depends on two parameters $p$ and $\rho$ \cite{footnote1}.
The $R_{N}(n)$ satisfy the normalization 
\beq
\sum_{n=0}^N\binom{N}{n}R_{N}(n)=1
\label{normrNn}
\eeq
and $\binom{N}{n}R_{N}(n)$ can be interpreted as the probability 
that the sum of $N$ Boolean random variables, correlated in a special way
(but remaining implicit), be exactly equal to $n$.
In this second example the starting point (\ref{defrNn})
is itself clearly motivated by
the formalism of $q$-statistical mechanics: the
last factor in the summand represents a so-called `$q$-product'
\cite{MTGM06} with $\rho$ in the role of
$1-q$; in the singular limit $\rho\to 0$ it reduces to $p^{-i}$. 

Moyano {\it et al.}~\cite{MTGM06} were again
interested in the large-$N$ limit with $n$ scaling as
$n=yN$, hence the limiting probability distribution
\beq
{\cal R}(y) = \lim_{N\to\infty} N \binom{N}{yN}R_{N}(yN),
\label{defcalR}
\eeq
where the first factor $N$ comes from the scaling between $n$ and $y$.
They evaluated the sum (\ref{defrNn}) 
numerically for typical values of $p$ and $\rho$ 
and for $N$ up to $1000$, 
taking extraordinary precautions to deal
with the cancellations caused by the alternating signs.
On the basis of their
numerical results Moyano {\it et al.}~\cite{MTGM06} conjecture  
that ${\cal R}(y)$ is a double-branched $q$-Gaussian, the two
branches joining in the center and having the same $q$ but
slightly different widths. After fitting they arrive at the relation
\beq
q = \frac{1-2\rho}{1-\rho}\,.
\label{relqrho2}
\eeq
The functional dependence of (\ref{relqrho2}) 
on $\rho$ bears resemblance to that
of $q$ in the first example, equation (\ref{relqrho1}).

For simplicity we will restrict ourselves in what follows 
to the case $p=\half$,
where some rewriting turns expression (\ref{defrNn}) into
\beq
R_{N}(n) = \frac{1}{ 2(1-2^{-\rho}) } \sum_{j=0}^n(-1)^{n-j}\binom{n}{j}
\frac {1} {[ N-j+(2^{\rho}-1)^{-1} ]^{\frac{1}{\rho}} }\,.
\label{rNnhalf}
\eeq 
Below we will show that it is possible, again, to evaluate this sum 
and find the limit function ${\cal R}(y)$ analytically. 


\subsection{The limiting law ${\cal R}(y)$}
\label{seclimitinglaw2}

Upon employing in (\ref{rNnhalf}) the integral representation
\beq
\frac{1}{X^k} = \frac{1}{\Gamma(k)} \int_0^\infty\!\dd\alpha\,
\alpha^{k-1}\ee^{-\alpha X}
\label{intrepr}
\eeq
with $X=N-j+(2^{\rho}-1)^{-1}$ and $k=1/\rho$ 
we find that the sum on $j$ represents the binomial expansion of 
$(\ee^\alpha-1)^n$. We get
\beq
R_{N}(n) = 
\frac{1}{A_\rho} \int_0^\infty\!\dd\alpha\,\alpha^{\frac{1-\rho}{\rho}}
\ee^{-(2^{\rho}-1)^{-1}\alpha - Nf(\alpha)}
\label{intexprrNn}
\eeq
with 
\bea
A_\rho &=& 2\Gamma(\tfrac{1}{\rho})(1-2^{-\rho})^{\frac{1}{\rho}},
\nonumber\\[2mm]
f(\alpha) &=& \alpha - y\log(\ee^\alpha-1).
\label{defAf}
\eea
The function $f(\alpha)$ has its maximum for 
$\alpha=\alpha^*\equiv\log(1- y)$.
In the large-$N$ limit the integral (\ref{intexprrNn}) 
is easily evaluated by the saddle
point method with the result
\bea
R_{N}(n) &=& \Big( \frac{2\pi}{Nf^{''}(\alpha^*)}\, \Big)^{\!\half}
           \ee^{-\alpha^*-Nf(\alpha^*)}\,\big[ 1\,+\,\ldots \big]
\nonumber\\[2mm]
&=& \Big( \frac{2\pi y(1-y)}{N} \Big)^{\!\half}
    \ee^{N[ y\log y + (1-y)\log(1-y)]}\,\big[ 1\,+\,\ldots \big],
\label{ressaddle}
\eea
where the dots indicate terms that vanish as $N\to\infty$ and 
we recall that $y=n/N$. 
This should now be combined with the large-$N$ expression 
of the binomial coefficient,
\beq
\binom{N}{n} = \big[ 2\pi N y(1-y) \big]^{-\half}\,
               \ee^{-N[ y\log y + (1-y)\log(1-y) ]}\,
               \big[\, 1 \,+\,\ldots\,\big],
\label{expbinom}
\eeq
which results from straightforward application of Stirling's formula.
Combining (\ref{defcalR}) with (\ref{ressaddle}) and (\ref{expbinom})
and setting
\beq
a_\rho = \frac{2-2^{\rho}}{2^{\rho}-1}
\label{defaQ}
\eeq
we obtain 
\beq
{\cal R}(y) = A_\rho^{-1}\,(1-y)^{a_\rho}\,
\big[ -\log(1- y) \big]^{\frac{1-\rho}{\rho}},
\qquad (0 \leq y\leq 1).
\label{rescalR}
\eeq
This is our result for the distribution ${\cal R}(y)$ studied by
Moyano {\it et al.}~\cite{MTGM06} in their special case $p=\half$.
Again, it is not the double-branched $q$-Gaussian 
that the authors \cite{MTGM06} conjectured it to be
(and which would have required it to be nonanalytic at its center). 

\begin{figure}
\begin{center}
\scalebox{.60}
{\includegraphics{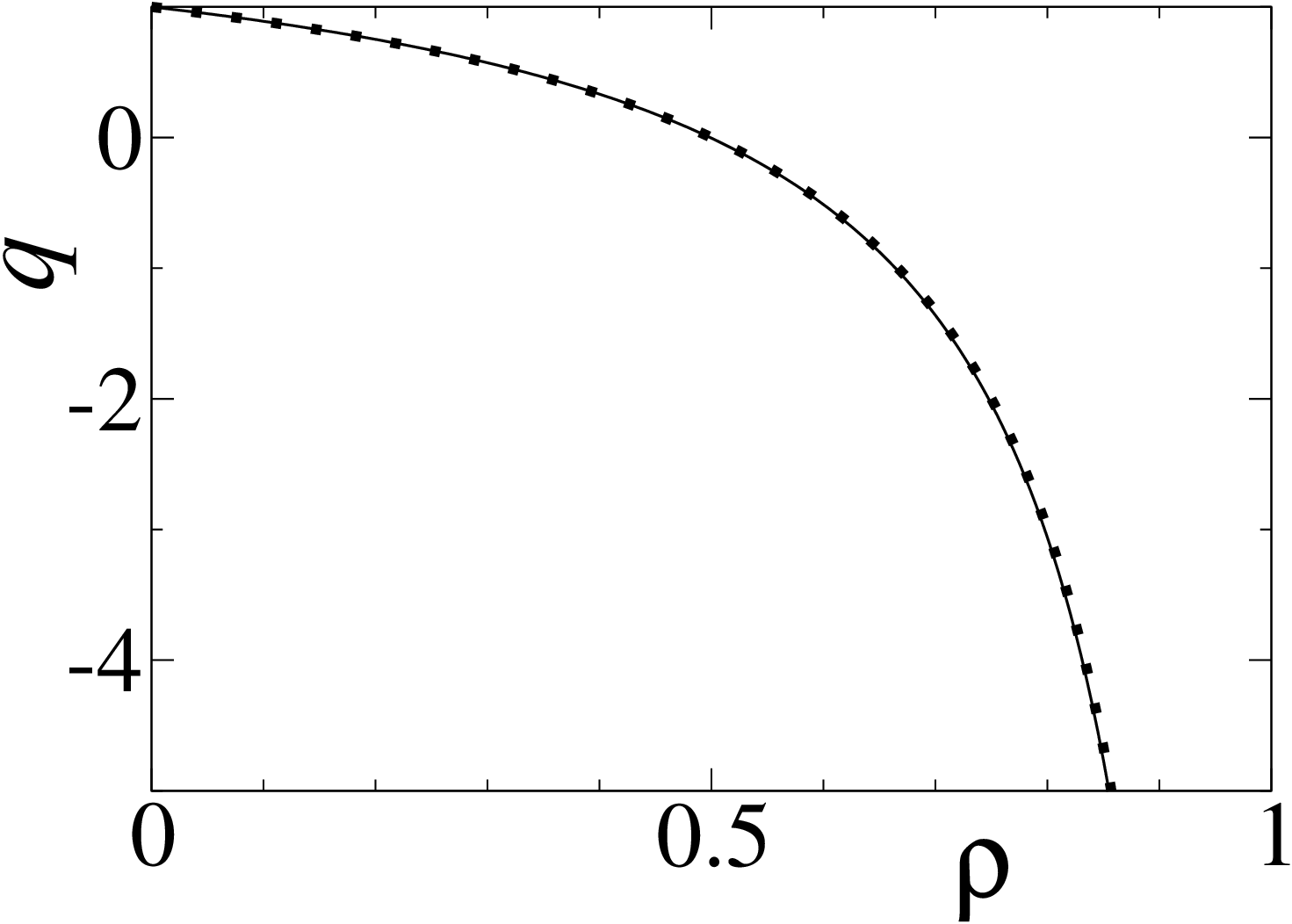}}
\end{center}
\caption{{\small  Comparison of two relationships between $q$ and $\rho$.
Solid line: equation (\ref{fitq2}) obtained analytically in this work.
Dotted line: equation (\ref{relqrho2})
conjectured by Moyano {\it et al.}~\cite{MTGM06} on the basis of numerical
work. The difference is hardly visible to the eye.}}
\label{figureqrho}
\end{figure}

\begin{figure}
\begin{center}
\scalebox{.60}
{\includegraphics{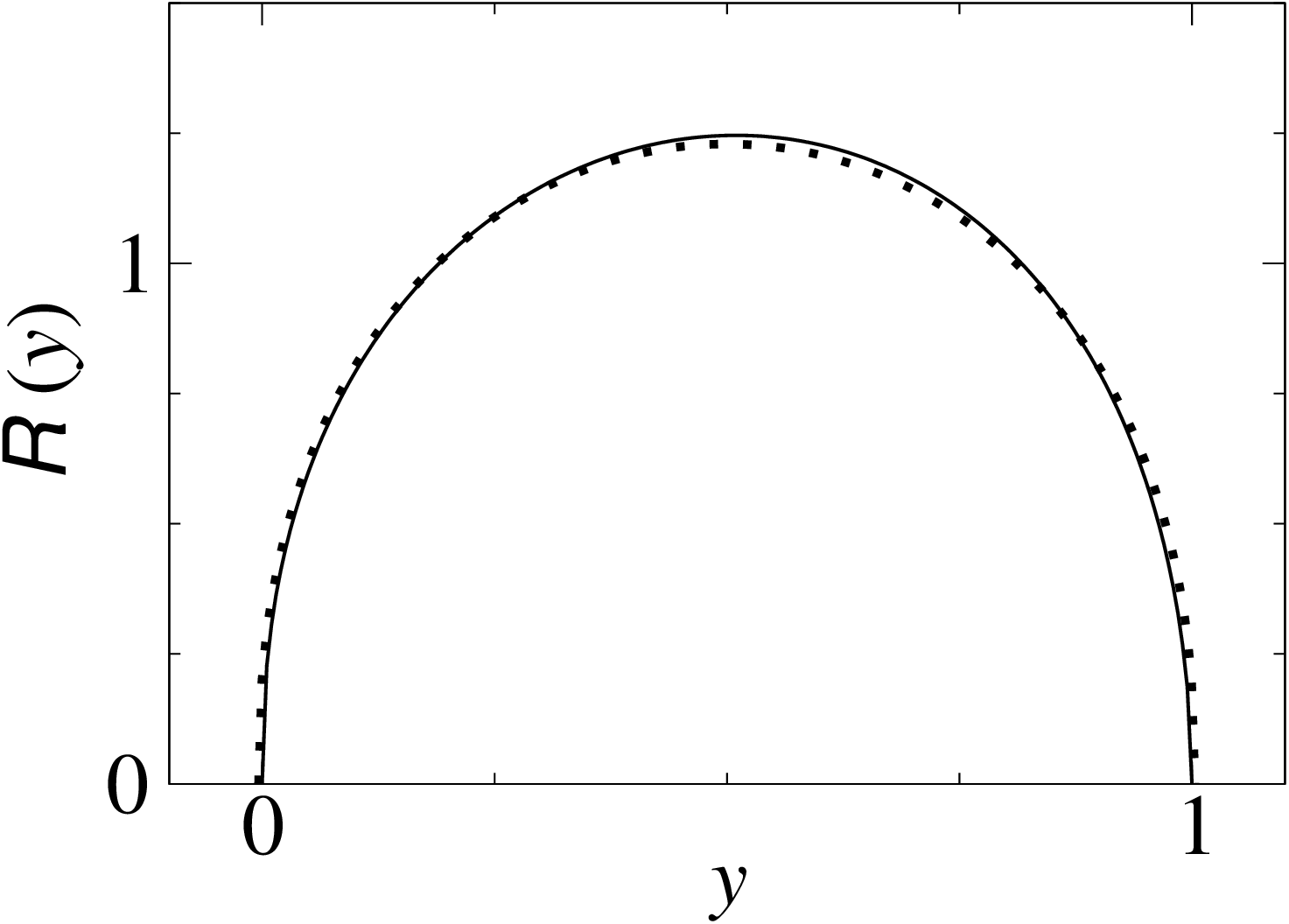}}
\end{center}
\caption{{\small  Solid line: the exact limit function ${\cal R}(y)$ 
given by equation (\ref{rescalR}) for the case $\rho=\tfrac{7}{10}$. 
Dotted line: the $q$-Gaussian 
approximation to this function for $q=-\tfrac{4}{3}$ (see text).
We improved the fit visually
by not centering the $q$-Gaussian at the exact maximum
$y^*_\rho$ but shifting it very slightly to the left.
The difference between the two curves is small but visible.}}
\label{figureRy}
\end{figure}


\subsection{Discussion of ${\cal R}(y)$}
\label{secdiscussion2}

The exponents $a_\rho$ and $(1-\rho)/\rho$ are both positive for $0<\rho<1$,
which is the interval of interest. Hence ${\cal R}(y)$ is bell-shaped
and vanishes in $y=0$ and $y=1$. 
One may verify explicitly the normalization 
$\int_{0}^1 {\cal R}(y)\dd y = 1$.
As had to be expected, ${\cal R}(y)$ has no point of symmetry.
It takes its maximum for $y=y^*_\rho$ given by
\beq
y^*_\rho = 1 \,-\,\exp\Big(\! -\frac{1-\rho}{\rho a_\rho} \Big).
\label{exprystar}
\eeq

We consider again two limiting cases of interest.
For $ \rho\to 0$ one shows after some algebra that ${\cal R}(y)$
approaches a Gaussian centered at $y=\half$ and of variance
$\rho(\log 2)^2$.
For $\rho\to 1$ we have $A_1=1$ and $a_1=0$, whence it follows that 
${\cal R}(y)$ is the block function ${\cal R}(y)=1$ on the interval 
$0\leq y\leq 1$. When $\rho\to 1$ the maximum $y^*_1$ of the distribution
has the nontrivial limit 
$y^*_1\equiv\lim_{\rho\to 1}y^*_\rho
= 1-\exp\big( -(2\log 2)^{-1} \big)=0.5139...$,
which is different from $\half$ in spite of ${\cal R}(y)$ being 
symmetric for $\rho=1$. 
In summary, for $0<\rho<1$, the sum ${\cal R}(y)$ varies between
a Gaussian and a block
function distribution, just as in the first example. 


\subsection{Best $q$-Gaussian fit to ${\cal R}(y)$}
\label{secbestfit2}

We will again find the $q$-Gaussian that best fits 
the exact result (\ref{rescalR}).
Upon expanding the latter around its maximum (\ref{exprystar}) one finds
\beq
{\cal R}(y)=\exp\Big(\! -\sum_{j=2}^\infty D_j(y-y^*_\rho)^j  \Big),
\label{expRy}
\eeq
which defines the coefficients $D_j$.
The best $q$ will be determined as a function of $\rho$ 
in the same way as above, that is, from the ratio of the
coefficients of the quadratic and the quartic terms according to 
$\half (1-q) = D_4/D_2^2$. This procedure `averages' over the asymmetry
due to the cubic and higher order odd terms in (\ref{expRy}) 
by simply neglecting them. 
After doing the algebra one finds \cite{footnote1}
\beq
q = 1 - \frac{ 11(1-\rho)^2 - 12(1-\rho)\rho a_\rho + 6\rho^2 a_\rho^2 }
{ 3(1-\rho)\rho a_\rho^2 }\,,
\label{fitq2}
\eeq
which has the small-$\rho$ expansion
\bea
q &=& 1 - \big[ \tfrac{11}{3}(\log 2)^2 - 4\log 2 + 2 \big]\rho 
- \big[11 (\log 2)^3 - \tfrac{29}{3}(\log 2)^2 + 2 \big] \rho^2 + \ldots
\nonumber\\[2mm]
&=& 1 - 0.98907...\rho - 1.01889...\rho^2 + \ldots\,.
\label{fitq2exp}
\eea
It appears from (\ref{fitq2}) 
that in this second example we do not reproduce the guess (\ref{relqrho2}) 
based by MTG \cite{MTGM06} on their numerical evaluation of ${\cal R}(y)$.
Nevertheless, 
the two relations between $q$ and $\rho$, 
equations (\ref{relqrho2}) and (\ref{fitq2}), 
are in very close {\it numerical\,} agreement, as shown 
in figure \ref{figureqrho}.
Concomitantly expansion (\ref{fitq2exp}), when
compared to the expansion $q=1-\rho-\rho^2+\ldots$ 
obtained from (\ref{relqrho2}),
shows a close numerical agreement between the coefficients.

In figure \ref{figureRy}, finally, we show
the limit function ${\cal R}(y)$  
for $\rho=\frac{7}{10}$ together with its $q$-Gaussian
approximation, which according to equation (\ref{relqrho2}) 
has $q=-\frac{4}{3}$.
No visible difference results if instead one were to take 
$q=-1.36074$ as would follow from (\ref{fitq2}).
Hence in this second example once more,
things conspire such that it becomes extremely difficult to distinguish
the true curve from its $q$-Gaussian approximant.


\section{Final remarks}
\label{secconclusion}

Probability distributions shaped as $q$-Gaussians,
if and when they occur,
are interpreted as a manifestation of
a $q$-generalized statistical mechanics \cite{Tsallis88}.
We have investigated the emergence of $q$-Gaussians as
probability distributions of {\it sums of strongly correlated variables.}
The possible role of $q$-Gaussians in a wider context has remained outside
the scope of this work.

The literature examples discussed in this note
illustrate the difficulty of identifying $q$-Gaussians.
Two cases where $q$-Gaussian distributions had been thought to occur,
turn out to correspond to analytically very different functions. 
The very close numerical agreement does not 
detract from this distinction of principle.
Although the first example does show a way of fine-tuning
correlations between random variables
such that their sum becomes $q$-Gaussian distributed,
this method works for any one out of a large class of distributions
and does not hint at a special role for $q$-Gaussians.
The probability laws constructed by this kind of fine-tuning are 
{\it limit distributions} in the plain sense that they appear 
when the number $N$ of terms in the sum tends to infinity.
They are not, however, or at least have not been shown to be, 
{\it attractors}, and this raises the question 
of why such finely adjusted correlations would occur in physics.
Efforts to let $q$-Gaussians play the role of attractors
within the framework of a $q$-generalized central limit theorem 
\cite{UTS06} have not yet led to examples that we can subject
to analysis.

The indeniably interesting properties of
$q$-Gaussians make it worthwile to investigate
the role they play in many-body systems in statistical
mechanics.  
However, from the examples studied in this work
we draw the conclusion that,
in the absence of convincing theoretical arguments,
one should exercise extreme caution
when interpretating non-Gaussian data 
in terms of $q$-Gaussians. 

\section*{Acknowledgments}
The authors have benefitted from discussions and correspondence 
with Professor Constantino Tsallis.
HJH thanks Professor Liacir S. Lucena for inviting him to participate
in the {\it Workshop on the Dynamics of Complex Systems}
(Natal, Brazil, 11-16 March 2007), where the idea for this work
originated.



\begin{thebibliography}{10}
\bibitem{footnote1}
In reference \cite{MTGM06}
our $\rho$ introduced in section \ref{secdefinition2} is called $1-q$
and our $q$ in section \ref{secbestfit2} is called $q_e$.
\end{thebibliography}
\end{document}